# INTEGRATED MOLAR ABSORPTIVITY OF MID- AND FAR-INFRARED SPECTRA OF GLYCINE AND OTHER SELECTED AMINO ACIDS

Susana Iglesias-Groth[1], Franco Cataldo[2(*)]

[1]*Instituto de Astrofisica de Canarias, Via Lactea s/n, 38200 La Laguna, Tenerife, Spain*
[2]*Actinium Chemical Research Institute, Via Casilina 1626A, 00133 Rome, Italy*

(*) corresponding author – e-mail: franco.cataldo@fastwebnet.it

**Abstract**

A selection of five proteinogenic amino acids, namely glycine, isoleucine, phenylalanine, tyrosine and tryptophan were studied in the mid-infrared and in the far-infrared with the purpose to facilitate the search and identification of these astrobiological and astrochemical relevant molecules in space environments. The molar extinction coefficients (ε) of all mid- and far-infrared bands were determined as well as the integrated molar absorptivities (ψ). The mid-infrared spectra of the five selected amino acids were recorded also at three different temperatures from -180°C to ambient temperature to +200°C. We measure the wavelength shift of the infrared bands caused by temperature and, for the most relevant or temperature-sensitive infrared bands, a series of linear equations were determined relating wavelength position with temperature. Such equations may provide estimates of the temperature of these molecules once detected in astrophysical objects and with the reported values of ε and ψ, it will be possible to estimate the relative abundance of these molecules in space environments.

## 1 INTRODUCTION

Proteinogenic and non-proteinogenic amino acids are known to be present in meteorites, especially in carbonaceous chondrites, since long time. An inventory of the amino acids found in meteorites was made by Commeyras et al. (2006), by Martins & Sephton (2009), Martins (2011), Pizzarello & Shock (2017) and Koga & Naraoka, (2017). It is now accepted the scenario that amino acids and other complex organic molecules which are prebiotic precursors of molecules and macromolecules of biological interest are formed in the icy mantles of dust grains and in the interstellar medium before the Solar System formation and these molecules are then trapped inside asteroids, comets and planetesimals which were delivered to the primordial Earth (Nuevo et al. 2007; Hudson et al. 2008; Kaiser et al. 2013; Kwok, 2012, 2016, 2019; Cataldo and Iglesias-Groth 2017; Cataldo, 2018). Of course, also thermal and aqueous alteration processes play a key role in the final amino acids distribution and their enantiomeric excess found in carbonaceous chondrites (Elsila et al. 2016). The experimental evidence of this scenario is becoming impressive through a number of experiments involving irradiation of simulated icy mantles of dust grains of the interstellar medium (see for example Bernstein et al. 2002; Caro et al. 2002; Hudson et al. 2008; Öberg, 2016; Modica et al. 2018; Kobayashi, 2019; Arumainayagam et al. 2019; Ciarravella et al. 2019) and astrophysical observations (Kwok, 2019; Scibelli, & Shirley, 2020). Furthermore, it has been shown that even elemental carbon either in the form of fullerene or under other allotropic states can be the source of atomic carbon suitable for the synthesis of amino acids under appropriate high energy fields (Iglesias-Groth et al. 2013; Ursini et al. 2019; Krasnokutski, Jäger, & Henning 2020). A series of studies have shown that all the proteinogenic amino acids and a selection of non-proteinogenic amino acids have a radiation stability sufficient to survive $4.6 \times 10^9$ years irradiation due to radionuclides decay, buried at a certain depth in asteroids and comets and shielded from the direct action of cosmic rays (Iglesias-Groth et al. 2011a: Cataldo, Ursini et al. 2011; Cataldo, Angelini et al. 2011; Cataldo, Ragni et al. 2011; Cataldo, Angelini et al. 2013; Cataldo, Iglesias-Groth et al. 2013; Cherubini et al. 2014). Thus, the scenario involving amino acids formation in the interstellar medium, followed by their incorporation in planetesimals, asteroids and comets in star-forming regions and late delivery to Earth is further reinforced by these radiation chemistry experiments performed on amino acids, following a calculation of the radiation dose received from radionuclides decay calculated many years ago by the Nobel laureate Harold Urey (1955; 1956). Very recently, also the possible formation of oligomers and peptides from amino acids under prebiotic conditions is receiving large attention (Kitadai and Nishiuki, 2019; Ligterink et al. 2019).

Another fascinating topic regarding prebiotic amino acids found in meteorites regards the enantiomeric excess systematically detected by many authors and pioneered by Pizzarello and Cronin (2000) and other authors. This topic has been comprehensively reviewed by Meierhenrich (2008) and more recently updated by Garcia et al. (2019). It is thought that the enantiomeric excess in initially racemic amino acids is induced by circularly polarized light irradiation from astronomical sources which may selectively transform a racemic mixture into a scalemic mixture (Bailey et al, 1998). This scenario is strongly supported by a series of interesting experimental results on the irradiation of interstellar ice analogues with circularly polarized UV light (Meierhenrich, 2008; Garcia et al. 2019).

With all these premises, an increased effort is put to the search of amino acids in space by astrophysical observations as briefly reviewed in our previous work (Iglesias-Groth & Cataldo, 2018). Similar to the case of fullerenes (e.g. Cami et al. 2010, Garcia-Hernandez et al. 2012; Iglesias-Groth 2019), amino acids could in principle be recognized in space through mid-infrared and also far-infrared spectroscopy, since new and forthcoming orbiting telescopes will cover also the far-infrared or terahertz spectral window with high spectral resolution and sensitivity. In particular, $C_{60}$ fullerene was discovered in the young planetary nebula TC1 and then in many other sources through Spitzer infrared orbiting telescope from its peculiar four band pattern in the infrared (Garcia-Hernandez et al. 2012). The integrated molar absorptivity of fullerene and fulleranes measured in laboratory (Iglesias-Groth et al. 2011b; Iglesias-Groth et al. 2012) were used for the estimation of the relative abundance of these molecules in certain planetary nebulae (Diaz-Luis et al. 2016). The $C_{60}$ molecule was not detected in the gas phase but adsorbed on solid surface, probably carbon dust (Cami et al. 2010; Garcia-Hernandez et al. 2012). By analogy, one assumes that if amino acids are embedded or generated within ices, they are more refractory than the other species, and thus perhaps as a young star forms or a comet dust sublimates, the refractory components such as amino acids would be left behind which may be identified at ambient, or even at moderately elevated temperatures through infrared spectroscopy.

Consequently, a selection of amino acids was spectroscopically studied in the mid and far-infrared (Iglesias-Groth & Cataldo 2018). As shown by Cami et al. (2010), Garcia-Hernandez et al. (2012) and others, for a quantitative analysis of the detected molecules in space the integrated molar absorptivity at least of the most relevant infrared bands is needed. Thus, the integrated molar absorptivity was determined in the spectra of fullerenes. The present work represents an extension of previous work on amino acid spectroscopy by Iglesias-Groth & Cataldo (2018) and integrated molar absorptivity of fullerenes (Iglesias-Groth et al. 2011b), fulleranes (Iglesias-Groth et al. 2012), and endohedral fullerenes (Garcia-Hernandez et al. 2020). Here, we report measurements of integrated molar absorptivity both in the mid-infrared and far-infrared spectral regions for a series of selected amino acids: glycine, isoleucine, phenylalanine, tyrosine and tryptophan.

It must be emphasized that our intention is to study all the proteinogenic amino acids and in this work we focused on the just mentioned five amino acids. While glycine is the simplest amino acid and was quite recently subjected to infrared studies in astrochemical relevant conditions (Maté et al. 2011; Souza-Corrêa, 2019), the integrated molar absorptivity of its mid and far-infrared bands are still unknown. Glycine, is the most interesting candidate to be searched in different astronomical objects. Isoleucine was little studied in the infrared, but it was found in meteorites analysis and hence is another interesting candidate to be searched in the future in different astronomical objects. Consequently, the infrared data of the present study are of certain relevance. Regarding the aromatic amino acids phenylalanine, tyrosine and tryptophan, it is true that they are seldom found in meteorites analysis, but we have included them in this work as a part of the planned systematic investigation both in the mid and in the far-infrared of all the proteinogenic amino acids.

## 2 EXPERIMENTAL

### 2.1 Materials

All the amino acids used in the present work were purchased from Aldrich-Merck. The spectroscopy matrices cesium iodide, low molecular weight polyethylene (PE), were also obtained from Aldrich-Merck.

### 2.2 – Laboratory equipment

The mid-infrared spectra were recorded in transmittance on Nicolet 6700 FT-IR spectrometer from Thermo-Scientific using a resolution of 4 $cm^{-1}$. The spectrometer could work at higher resolution, but 4 $cm^{-1}$ is the optimal condition for samples in the solid state embedded in a matrix like CsI, for instance. The low-temperature apparatus consisted of a variable temperature cell from Specac model P/N 21525 equipped with heated KBr windows and a temperature-controlled sample holder, which is able to work in the range between +250°C to -180°C. The low temperature limit of -180°C was reached using liquid nitrogen. The variable temperature cell was evacuated with a Buchi vacuum pump model V-710 equipped with four diaphragm heads and a three-stage vacuum creation process, which delivers 3.1 $m^3$ $h^{-1}$ and an absolute vacuum of 1 mbar.

The far-infrared spectra from 600 $cm^{-1}$ down to 50 $cm^{-1}$ were recorded with the Nicolet 6700 FT-IR spectrometer always at a resolution of 4 $cm^{-1}$, after opportune adjustments of the optical bench (change of the beam splitter and the detector). Furthermore, the entire spectrometer optical bench and the sample compartment was continuously purged with completely dry air free from $CO_2$ at 12 L/min. The CsI, or PE pellet used in the infrared spectra measurement was produced by a Silfradent double column press equipped with a pressure meter. Aldrich macro-micro pellet die was loaded with the pre-weighed CsI and amino acid, closed, compressed in double column press while evacuated at 2 mbar. The thickness and the diameter of the pellets was measured with a Somet digital micrometer having a sensitivity of 0.01 mm.

We shall note that transitions with very high asorbance may be subject to deviations from Beer-Lambert law and those band strengths could be infraestimated.

**2.3 – Pellet preparation for the determination of the integrated molar absorptivity and the molar extinction coefficient in the mid infrared.**

In a typical procedure, the selected amino acid (1.2 mg) was quickly mixed with 250.0 mg of CsI in an agate mortar and the two components were finely ground and mixed together. The powder was transferred into the macro–micro pellet die, and compressed at 6 tonnes cm$^{-2}$ with the Silfradent press. The resulting pellet, with a measured thickness of 0.71 mm, was mounted into the sample holder of the Specac variable temperature cell and inserted into the cell. The cell was then evacuated with the aid of the Buchi pump to a vacuum of 0.1 torr and then heated gradually at +60°C in order to permit the humidity, which is eventually absorbed on the internal surfaces of the cell and in the KBr pellet, to evaporate. In order to go below room temperature, use was made of liquid nitrogen, added cautiously and in small amounts in the cavity present inside the Specac cell. Such a cavity is connected with the sample holder and permits us to cool the sample to the desired temperature. The temperature of the sample was monitored with thermocouples present in the cell. The lowest temperature reached with this apparatus was −180°C, while the highest temperature was +250°C. Heating is provided by the Joule effect supplied to the Specac cell by an external thermal control unit. A more detailed description of the apparatus (photographs) used for the measurement can be found in Cataldo & Iglesias-Groth (2010). The spectra were recorded in absorbance mode. The measurement of the intensity (height) of the absorption bands with automatic subtraction of the baseline was made through the OMNIC software from Thermo, dedicated to the FTIR spectrometer. Similarly, the integrated band intensity was also measured through the OMNIC software. The error level both in the determination of ε and ψ values is in the range of ±13%.

**2.4 - Pellet preparation for the determination of the integrated molar absorptivity and the molar extinction coefficient in the far-infrared.**

The selected amino acids (1.2 mg) was mixed in an agate mortar with polyethylene (220 mg), a matrix suitable for far infrared spectroscopy. The pellet was compressed in the Silfradent press at 6 tonn/cm$^2$ inside the macro–micro pellet die. The resulting pellet was obtained with a measured diameter of 12.55 mm and a measured thickness 0.73 mm. The far infrared spectra were collected exclusively at room temperature in absorbance mode without the use of the Specac cell. Also in the far infrared the error estimation in both of ε and ψ values is in the range of ±13%.

## 3. RESULTS AND DISCUSSION

### 3.1 – On the molar extinction coefficients and integrated molar absorptivity of the mid-infrared bands of selected amino acids

The detailed explanation on the procedure for the determination of the molar extinction coefficients (ε) and the integrated molar absorptivity (ψ) of the infrared bands can be found in a series of papers on fullerenes, fulleranes (the hydrogenated fullerenes) and endohedral fullerenes infrared spectroscopy (Cataldo & Iglesias-Groth, 2010; Iglesias-Groth et al. 2011b and 2012; Garcia-Hernandez et al. 2020). It is worth to remind here that the knowledge of these parameters are important in astrochemistry for the quantitative estimation of the abundance of a given molecule once it has been identified qualitatively in a certain space environment through the infrared band pattern. A typical example of the application of the integrated molar absorptivity is reported in the work of Cami et al. (2010), where the ψ parameter of the fullerene infrared bands was employed in the evaluation of the relative abundance of $C_{60}$ fullerene present in the young planetary nebula TC-1. Similarly, Diaz-Luis et al. (2016) employed the ψ parameter determined for fullerenes and fulleranes (Iglesias-Groth et al. 2011b and 2012) for the estimation of the relative abundance of these molecules in a series of planetary nebulae.

Through the Lambert-Beer law, the molar extinction coefficient (or molar attenuation coefficient) is expressed as follows:

$$\varepsilon = A\,(bc)^{-1} \quad\quad\quad [1]$$

i.e. for a selected wavelength λ or (wavenumber $\nu = \lambda^{-1}$) the corresponding absorbance A in the spectrum multiplied by the path length b of the matrix containing the sample expressed in cm and the concentration of the sample under analysis in the matrix expressed in mol•L$^{-1}$, it gives the molar extinction coefficient $\varepsilon_\lambda$ expressed in L•mol$^{-1}$•cm$^{-1}$ (Cataldo & Iglesias-Groth, 2010; Iglesias-Groth et al. 2011b and 2012).

On the other hand, the integrated molar absorptivity ψ is derived by the integration of the absorbance over the entire absorption band:

$$\psi = \int \varepsilon\, d\lambda = (bc)^{-1} \int A\, d\lambda \quad\quad\quad [2]$$

The integrated absorptivity in an infrared spectrum having wavenumbers in the abscissa has the dimensions of $cm^{-1}$. If also the path length b is expressed in cm and the concentration is expressed in $mol \cdot cm^{-3}$, then ψ has the dimensions of $cm \cdot mol^{-1}$. By multiplying by a factor $10^{-5}$ it is possible to convert into the astrochemical practical value of $km \cdot mol^{-1}$ (Cataldo & Iglesias-Groth, 2010; Iglesias-Groth et al. 2011b and 2012).

The mid infrared spectra of proteinogenic α-amino acids and the assignment of the vibrations can be found in numerous work and in reviews, for example those from Parker (1971) and from Barth (2000). On the other hand, little is known about the molar extinction coefficients (ε) and moreover about the integrated molar absorptivity (ψ) of the amino acids infrared bands. Regarding the molar absorption coefficients of amino acids, a reference work is certainly that from Wolpert, & Hellwig (2006), where the ε values of all proteinogenic amino acids were determined in aqueous solutions. However, the ε and moreover ψ values of α-amino acids in the solid state, which are of more relevance in astrochemistry and astrobiology, have not been determined yet. In the present work, a selection of amino acids, starting from the simplest glycine and involving also phenylalanine, tyrosine, isoleucine and tryptophan were studied in the solid state for the determination of ε and ψ values of the most relevant infrared bands.

Tables 1-5 show the results of our determination of both ε and ψ values on glycine, isoleucine, phenylalanine, tyrosine and tryptophan respectively in the solid state. The data are derived from the spectra reported in Fig. 1-6. Regarding the integrated molar absorptivity, the Tables 1-5 report also the integration ranges over which, the ψ value of each relevant band was determined.

It is important to warn here that the reported molar absorptivities have been obtained under specific laboratory conditions and should be used with caution when attempting abundance determinations of amino acids in space. They could be useful to provide insight on the relative abundances of these species when observed in the same spectral regions and to obtain order of magnitude estimates for abundances. Precise abundance determinations will require that absorptivity values of amino acids are measured in laboratory experiments resembling as much as possible the conditions of interstellar space.

It is well known that in aqueous solutions the position of the infrared bands of amino acids are greatly affected by the pH of the solution or, which is the same, the degree of dissociation and protonation of the amino acids. A similar destiny is reserved also to the ε values of the amino acids, i.e. they are affected by the pH of the solution (Wolpert & Hellwig, 2006). Indeed, a direct comparison of the amino acids ε values determined in the solid state and in aqueous solution leads to large differences.

However, for glycine, the infrared bands at 9.67, 10.99 11.20 μm (1034, 910 and 893 cm$^{-1}$) show the same ε values both in the solid state and in aqueous solution as reported by Wolpert & Hellwig, (2006). These bands are due to $\nu_{CN}$ and $\nu_{CC}$ stretching modes which evidently are not too much affected by the environment where the glycine is surrounded. Furthermore, the good match between the ε value determined in CsI matrix and in aqueous solution is another confirmation of the goodness of the collected data. The good match in molar extinction coefficients either determined in the solid state (CsI) or in aqueous solution as reported by Wolpert & Hellwig, (2006) can be observed in the case of isoleucine bands at 7.06, 8.87, 10.07 and 10.38 μm (1417, 1128, 993 and 964 cm$^{-1}$), phenylalanine bands at 6.86 and 8.15 μm (1458 and 1226 cm$^{-1}$), tyrosine bands at 6.22 and 9.10 μm (1608 and 1099 cm$^{-1}$) as well as in the case of the tryptophan bands at 6.00, 8.12 and 17.00 μm (1666, 1231 and 588 cm$^{-1}$). In all these cases the same comments as expressed for glycine applies for these other amino acids. The infrared bands displaying similar ε values in the solid state and in aqueous solutions are necessarily due to stretching or bending modes not directly affected by the interaction of the -COO$^-$ group and -NH$_2$ moiety of amino acids with aqueous solutions.

## 3.2– The molar extinction coefficients and integrated molar absorptivity of the far-infrared bands of selected amino acids

As discussed in a previous work (Iglesias-Groth & Cataldo, 2018) the determination of the far infrared spectra of amino acids is of paramount importance for the future search of these molecules in space since the new orbiting telescopes will be able to cover this spectral range. Indeed, the far-infrared spectra of all the proteinogenic amino acids were recently determined and the far-infrared vibrations assigned and discussed in a broader perspective involving also previous literature data (Iglesias-Groth & Cataldo, 2018). The amino acids vibrations in the far infrared spectral range are essentially due to skeletal deformations, torsions, bending, rocking and wagging modes. Furthermore, hydrogen bonding modes can contribute in this spectral range.

However, also in the case of far infrared spectra, it is important to know the molar extinction coefficients (ε) and the integrated molar absorptivities (ψ) of the far infrared bands. In fact, if any amino acid bands will be ever detected in the far infrared in some astrophysical environment or object, through the known ε and ψ values it will be possible to determine the relative abundance.

The far infrared spectral range is nowadays preferably referred as terahertz spectral range. To cover this low frequency and low energy spectral range, it is necessary to change the beam splitter of the FT-IR spectrometer. The standard beam splitter to cover the mid-infrared spectral range is made in KBr but to reach the far infrared the most common beam splitter is made by Mylar (Cataldo et al.

2019). In our Nicolet 6700 spectrometer, not only the beam splitter but also a specific far infrared detector should be mounted substituting the standard detector for the mid-infrared. With these changes in the optical bench, our FT-IR spectrometer can cover the typical far-infrared spectral window comprised between 16.67 and 200 μm (600 and 50 cm$^{-1}$). Furthermore, instead of CsI (which in any case may be satisfactory in the far infrared spectral range) use was made of polyethylene as pellet matrix. The ε and ψ values in the far-infrared of each selected amino acids are reported in bold italics character at the bottom of each Table 1-5 and are derived from Fig.6. In all data reported in Tables 1-5 the tail of the mid-infrared bands overlap with the head of the far-infrared spectral window so that certain infrared bands were measured two times with mid-infrared spectrometer and sample configuration (Fig. 1-5) and with far-infrared spectrometer and sample configuration (Fig.6). It is possible to observe a fair agreement in the ε and ψ values determined in the different conditions and this is a further grant of the goodness of the values determined.

**3.3– Infrared spectra from -180°C to room temperature to +200°C**

The mid-infrared spectra of the selected amino acids were studied in a wide range of temperature comprised between -180°C (i.e. the liquid nitrogen temperature), at ambient temperature and then heated to the maximum permitted temperature before the decomposition. In Fig. 1 are reported the infrared spectra of glycine at -180°C, +18°C and +170°C. The latter temperature was considered the highest reachable temperature without glycine decomposition. The low temperature infrared spectroscopy of glycine was studied recently (Chernobai et al. 2007; Chasalov et al. 2008 and ref. cited therein) and the spectral changes observed are due to bond force constant dependence from temperature as well as polymorphism of glycine crystals. We will not enter into a detailed analysis of the vibrational assignments and discussion of the crystalline state transition of glycine since a complete work can be found in Chernobai (2007). Instead, in an astrochemical perspective, it is necessary to know how the infrared spectrum of any given amino acid changes as function of temperature. For example, in Fig. 1 it is shown that the glycine infrared spectrum at -180°C permits to resolve the broad band at 19.85 μm at +18°C into two well resolved bands at 19.73 and 18.32 μm. Furthermore, all the infrared bands of glycine appear narrower and sharper at low temperature and certain doublets like of instance those at 10.86 and 11.15 μm or that at 8.77 and 8.95 or even that at 6.21 and 6.29 μm appear much more resolved at -180°C. In case glycine (or any other amino acid) will be firmly discovered in space, it is possible to advocate, that the actual temperature of glycine (or any other amino acid) could be estimated from certain details of the detected infrared spectrum. Such details may indeed involve the fact that a band like that at 19.85 μm is resolved into two bands, each band width of the spectrum, the degree of resolution of selected doublets and the relative intensity ratio as well as the possibility to detect or not certain weak bands such as those at 6.21, 6.92 and the shoulder at 7.63 μm. Another spectral tool to estimate the actual temperature of the detected glycine in space could be achieved through the use of linear equations describing the shift of the band position as function of temperature. In general, the infrared band are shifted to shorter wavelengths (and higher frequencies) at lower temperatures. This phenomenon called "temperature effect" is due to a combination of different effects, involving quantum mechanical and changes in bond length (and hence force constant) as reviewed by Avram (1972). By studying the most sensitive infrared bands of glycine as function of temperature (Fig. 1), linear equations were derived and reported in Table 1. By determining the exact position of the glycine infrared bands at 8.83, 10.99, 14.32 and 16.45 μm in an eventual infrared spectrum in space, it is possible to estimate the actual temperature of the molecules. Thus, for example taking the following equation from Table 1 (with T absolute temperature):

$$\lambda_{8.83} = 2.900 \times 10^{-4} \, T + 8.743 \qquad [3]$$

then,

$$T = (\lambda_{8.83} - 8.743) / 2.900 \times 10^{-4} \qquad [4]$$

The intercept in eq. [3] is found at 8.743 μm and it represents the position expected for this absorption band at nearly 0 K (Iglesias-Groth et al. 2012). Thus, for all equations showing wavelength dependence from temperature reported in Table 1, from the intercept it is possible to predict the given band position at nearly 0 K.

The isoleucine mid-infrared spectra in a wide range of temperatures are reported in Fig.2. The full analysis of the vibrational bands of isoleucine was reported for example by Moorthi et al (2014). Similar comments as already detailed for glycine spectra can be applied as well to isoleucine spectra, especially to that at -180°C which provides the better resolution spectrum. Table 2 reports also the linear equations determined for all the isoleucine infrared bands which show the higher temperature dependence or the infrared bands with the higher ε or ψ values, which should be those more easily detectable in a space environment. These bands could be used to estimate the temperature of isoleucine in a space environment.

A full account of phenyalanine vibrational analysis can be found for example in the work of Kaczor et al. (2004). The relative infrared spectra of phenylalanine are reported in Fig. 3 recorded at -180°C, +20°C and +205°C. The infrared bands of phenylalanine are less sensitive to temperature and undergo smaller bathochromic shifts from low temperature to higher temperatures in comparison to the case of glycine and isoleucine, judging from the slopes of the linear equations connecting wavelengths with temperature (see Table 3).

Tyrosine is molecularly a strict parent of phenylalanine and the infrared spectroscopical review of these two molecules can be found in the Barth's reviews (Barth 2000; 2007). The infrared spectra of tyrosine at three different temperatures can be found in Fig. 4, while Table 4 reports also wavelength dependence from temperature of the most relevant infrared bands. The latter properties are completely equivalents to those measured in the case of phenylalanine.

Fig. 5 shows the infrared spectra at -180°C, +20°C and +200°C of tryptophan and also in this case the low temperature spectrum is definitely richer of better resolved infrared bands than the higher temperature spectra. Table 5 shows that the temperature dependence of certain infrared bands of tryptophan is quite high. This is especially true in the case of the band at 13.43 μm, with the highest temperature dependence among the five amino acids studied in the present work. The vibrational analysis of tryptophan in the mid-infrared was again summarized by Barth (2000; 2007).

Fig. 6 shows the far-infrared spectra of the selected amino acids recorded at ambient temperature. The main bands position is summarized at the bottom of Tables 1-5 in bold italics character. Due to limitations in our experimental set-up, it was possible to collect the far-infrared spectra only at ambient temperature and not at the extreme temperatures as instead was made in the mid-infrared spectral window. A discussion of the far infrared spectra of amino acids can be found in the previous work of Iglesias-Groth and Cataldo (2018).

**CONCLUSIONS**

The results of this work are a series of key tool for searching five selected amino acids (glycine, isoleucine, phenylalanine, tyrosine and tryptophan) in space. First of all, the mid-infrared spectra of the five amino acids are reported, recorded at three different temperatures i.e. -180°C, +20°C and +200°C. Furthermore, also the far-infrared spectra of the selected amino acids are reported as well at +20°C. Both the integrated molar absorptivity ($\psi$) and the molar extinction coefficients ($\varepsilon$) of all mid- and far-infrared bands were determined and are reported. With these data it shall be possible not only to search and eventually identify the selected amino acids in different space environments, but will allow also the estimation of the local abundance of the eventually detected amino acids through the use of the $\psi$ and $\varepsilon$ values. The dependence from temperature of certain key bands (the most intense or the most temperature-dependent) of the selected amino acids, may allow to estimate the actual temperature of the amino acids once eventually discovered in a given astrophysical object.

**ORCID**

Franco Cataldo http://orcid.org/0000-0003-2607-4414

**REFERENCES**

Arumainayagam, C. R., Garrod, R. T., Boyer, M. C., Hay, A. K., Bao, S. T., Campbell, J. S., Wang, J., Nowak, C.M., Arumainayagam, M. R., Hodge, P. J. (2019). Extraterrestrial prebiotic molecules: photochemistry vs. radiation chemistry of interstellar ices. *Chem Soc Rev* 48: 2293-2314.

Avram, M., Mateescu G.D. (1972) *Infrared Spectroscopy. Applications in Organic Chemistry*. Wiley-Interscience, New York, p.52-53.

Bailey J., Chrysostomou, A., Hough, J.H., Gledhill, T.M., McCall, A., Clark, S., Ménard, F., Tamura, M. (1998) Circular polarization in star-formation regions: Implications for biomolecular homochirality. *Science* 281: 672–674.

Barth, A. (2000) The infrared absorption of amino acid side chains. *Progr Biophys Molec Biol* 74: 141-175.

Barth, A. (2007) Infrared spectroscopy of proteins. *Biochim Biophys Acta (BBA)-Bioener* 1767: 1073-1101.

Bernstein, M. P., Dworkin, J. P., Sandford, S. A., Cooper, G. W., Allamandola, L. J. (2002) Racemic amino acids from the ultraviolet photolysis of interstellar ice analogues. *Nature* 416: 401-403.

Cami, J., Bernard-Salas, J., Peeters, E., Malek, S. E. (2010) Detection of $C_{60}$ and $C_{70}$ in a young planetary nebula. *Science* 329: 1180-1182.

Caro, G. M., Meierhenrich, U. J., Schutte, W. A., Barbier, B., Segovia, A. A., Rosenbauer, H., Greenberg, J. M. (2002) Amino acids from ultraviolet irradiation of interstellar ice analogues. *Nature* 416: 403-406.

Cataldo, F., Iglesias-Groth, S. (Eds.), (2010) *Fulleranes: The Hydrogenated Fullerenes.* Springer Science & Business Media, Dordrecht, NL.

Cataldo, F., Ursini, O., Angelini, G., Iglesias-Groth, S., Manchado, A. (2011) Radiolysis and radioracemization of 20 amino acids from the beginning of the Solar System. *Rendiconti Lincei*, 22: 81-94.

Cataldo, F., Angelini, G., Iglesias-Groth, S., Manchado, A. (2011) Solid state radiolysis of amino acids in an astrochemical perspective. *Radiat Phys Chem* 80: 57-65.

Cataldo, F., Ragni, P., Iglesias-Groth, S., Manchado, A. (2011) Solid state radiolysis of sulphur-containing amino acids: cysteine, cystine and methionine. *J Radioanal Nucl Chem* 287: 573-580.

Cataldo, F., Angelini, G., Hafez, Y., Iglesias-Groth, S. (2013) Solid state radiolysis of non-proteinaceous amino acids in vacuum: astrochemical implications. *J Radioanal Nucl Chem* 295: 1235-1243.

Cataldo, F., Iglesias-Groth, S., Angelini, G., Hafez, Y. (2013) Stability toward high energy radiation of non-proteinogenic amino acids: implications for the origins of life. *Life*, 3: 449-473.

Cataldo, F., Iglesias-Groth, S. (2017) Radiation chemical aspects of the origins of life. *J Radioanal Nucl Chem* 311: 1081-1097.

Cataldo, F. (2018) Radiolysis and radioracemization of RNA ribonucleosides: implications for the origins of life. *J Radioanal Nucl Chem* 318: 1649-1661.

Chernobai, G. B., Chesalov, Y. A., Burgina, E. B., Drebushchak, T. N., Boldyreva, E. V. (2007) Temperature effects on the IR spectra of crystalliine amino acids, dipeptides, and polyamino acids. I. Glycine. *J Struct Chem* 48: 332-339.

Chesalov, Y. A., Chernobai, G. B., Boldyreva, E. V. (2008) Study of the temperature effect on IR spectra of crystalline amino acids, dipeptids, and polyamino acids. III. α-Glycylglycine. *J Struct Chem* 49: 1012-1014.

Cherubini, C., Ursini, O., Cataldo, F., Iglesias-Groth, S., Crestoni, M. E. (2014) Mass spectrometric analysis of selected radiolyzed amino acids in an astrochemical context. *J Radioanal Nucl Chem* 300: 1061-1073.

Ciaravella, A., Jiménez-Escobar, A., Cecchi-Pestellini, C., Huang, C. H., Sie, N. E., Caro, G. M., Chen, Y. J. (2019). Synthesis of complex organic molecules in soft X-ray irradiated ices. *Astrophys J* 879: 21.

Commeyras A, Boiteau L, Vandenabeele-Trambouze O, Selsis F (2006) In: Gargaud M, Barbier B, Martin H, Reisse J (eds) *Lectures in Astrobiology. Part 2: from Prebiotic Chemistry to the Origin of Life on Earth*, vol 1. Springer, Berlin.

Diaz-Luis, J.J., Garcia-Hernandez, D.A., Manchado, A. Cataldo, F. (2016). A search for hydrogenated fullerenes in fullerene-containing planetary nebulae. *Astron. Astrophys.* 589: A5 (7 pages).

Elsila, J. E., Aponte, J. C., Blackmond, D. G., Burton, A. S., Dworkin, J. P., Glavin, D. P. (2016). Meteoritic amino acids: Diversity in compositions reflects parent body histories. *ACS Central Sci* 2: 370-379.

Garcia, A. D., Meinert, C., Sugahara, H., Jones, N. C., Hoffmann, S. V., Meierhenrich, U. J. (2019) The astrophysical formation of asymmetric molecules and the emergence of a chiral bias. *Life* 9: 29.

García-Hernández, D. A., Villaver, E., García-Lario, P., Acosta-Pulido, J. A., Manchado, A., Stanghellini, L., Shaw, R.A., Cataldo, F. (2012). Infrared study of fullerene planetary nebulae. *Astrophys. J.* 760: 107.

García-Hernández, D. A., Manchado, A., Cataldo, F. (2020). [Li@ $C_{60}$] $PF_6$: Infrared spectra from 90K to 523K; Determination of the molar extinction coefficients and integrated molar absorptivity. *Fullerenes Nanot Carbon Nanostruct* 28:474-479

Hudson, R. L., Moore, M. H., Dworkin, J. P., Martin, M. P., Pozun, Z. D. (2008) Amino acids from ion-irradiated nitrile-containing ices. *Astrobiol* 8: 771-779.

Iglesias-Groth, S., Cataldo, F., Ursini, O., Manchado, A. (2011a) Amino acids in comets and meteorites: stability under gamma radiation and preservation of the enantiomeric excess. *Monthly Notices Royal Astronom Soc* 410: 1447-1453.

Iglesias-Groth, S., Cataldo, F., Manchado, A. (2011b) Infrared spectroscopy and integrated molar absorptivity of $C_{60}$ and $C_{70}$ fullerenes at extreme temperatures. *Monthly Notices of the Royal Astronom Soc* 413: 213-222.

Iglesias-Groth, S., García-Hernández, D. A., Cataldo, F., Manchado, A. (2012) Infrared spectroscopy of hydrogenated fullerenes (fulleranes) at extreme temperatures. *Monthly Notices Royal Astronom Soc* 423: 2868-2878.

Iglesias-Groth, S., Hafez, Y., Angelini, G., Cataldo, F. (2013) γ Radiolysis of $C_{60}$ fullerene in water and water/ammonia mixtures: relevance of fullerene fate in ices of interstellar medium. *J Radioanal Nucl Chem* 298: 1073-1083.

Iglesias-Groth, S., Cataldo, F. (2018) Far-infrared spectroscopy of proteinogenic and other less common amino acids. *Monthly Notices Royal Astronom Soc* 478: 3430-3437.

Iglesias-Groth, S. (2019) Fullerenes in the IC 348 Star cluster of the Perseus molecular cloud. *Monthly Notices Royal Astronom Soc* 489: 1509-1518.

Kaczor, A., Reva, I. D., Proniewicz, L. M., Fausto, R. (2006) Importance of entropy in the conformational equilibrium of phenylalanine: a matrix-isolation infrared spectroscopy and density functional theory study. *J Phys Chem A* 110: 2360-2370.

Kaiser, R. I., Stockton, A. M., Kim, Y. S., Jensen, E. C., Mathies, R. A. (2013). On the formation of dipeptides in interstellar model ices. *Astrophys. J* 765: 111.

Kitadai, N., Nishiuchi, K. (2019) Thermodynamic impact of mineral surfaces on amino acid polymerization: aspartate dimerization on goethite. *Astrobiol* 19: 1363-1376.

Kobayashi, K. (2019) Prebiotic synthesis of bioorganic compounds by simulation experiments. In *Astrobiology* edited by Yamagishi, A., Kakegawa, A., Usui, T. Springer, Singapore, pp. 43-61.

Koga, T., Naraoka, H. (2017) A new family of extraterrestrial amino acids in the Murchison meteorite. *Scientific Reports 7*: 636 (8 pages).

Krasnokutski, S. A., Jäger, C., Henning, T. (2020) Condensation of atomic carbon: possible routes toward glycine. *Astrophys J* 889: 67.

Kwok, S. (2012) *Organic Matter in the Universe*. Wiley-VCH, Weinheim.

Kwok, S. (2016) Complex organics in space from Solar System to distant galaxies. *Astron Astrophys Rev* 24: 8.

Kwok, S. (2019) Formation and delivery of complex organic molecules to the solar system and early earth. in *Handbook of Astrobiology* edited by Kolb, V.M., Chapter 4.2, p. 165-173, CRC Press: Boca Raton.

Ligterink, N. F. W., Terwisscha van Scheltinga, J., Taquet, V., Jørgensen, J. K., Cazaux, S., van Dishoeck, E. F., Linnartz, H. (2018) The formation of peptide-like molecules on interstellar dust grains. *Monthly Notices of the Royal Astronom Soc* 480: 3628-3643.

Martins Z, Sephton M.A. (2009) Extraterrestrial Amino Acids. In: Hughes A.B. (ed) *Amino acids, Peptides and Proteins in Organic Chemistry. Origins and Synthesis of Amino Acids*, Vol.1, Wiley-VCH, Weinheim, pp.3-42.

Martins, Z. (2011) Organic chemistry of carbonaceous meteorites. *Elements* 7: 35-40.

Maté, B., Rodriguez-Lazcano, Y., Galvez, O., Tanarro, I., Escribano, R. (2011). An infrared study of solid glycine in environments of astrophysical relevance. *Phys. Chem. Chem. Phys.* 13: 12268-12276.

Meierhenrich, U. (2008) *Amino Acids and the Asymmetry of Life: Caught in the Act of Formation*. Springer Science, Berlin.

Modica, P., Martins, Z., Meinert, C., Zanda, B., d'Hendecourt, L. L. S. (2018) The amino acid distribution in laboratory analogs of extraterrestrial organic matter: a comparison to CM chondrites. *Astrophys J* 865: 41.

Moorthi, P. P., Gunasekaran, S., Ramkumaar, G. R. (2014) Vibrational spectroscopic studies of Isoleucine by quantum chemical calculations. *Spectrochim Acta Part A: Molec Biomolec Spectr* 124: 365-374.

Nuevo, M., Chen, Y. J., Yih, T. S., Ip, W. H., Fung, H. S., Cheng, C. Y., Tsai, H.r., Wu, C. Y. (2007). Amino acids formed from the UV/EUV irradiation of inorganic ices of astrophysical interest. *Adv Space Res* 40: 1628-1633.

Öberg, K. I. (2016) Photochemistry and astrochemistry: Photochemical pathways to interstellar complex organic molecules. *Chem Rev* 116: 9631-9663.

Parker, F. S. (1971) Amino Acids, Related Compounds, and Peptides. In *Applications of Infrared Spectroscopy in Biochemistry, Biology, and Medicine* (pp. 173-187). Springer, Boston, MA.

Pizzarello S, Cronin JR (2000) Non-racemic amino acids in the Murray and Murchison meteorites. *Geochim Cosmochim Acta* 64:329–338

Pizzarello, S., Shock, E. (2017) Carbonaceous chondrite meteorites: The chronicle of a potential evolutionary path between stars and life. *Origins of Life Evol Biosph* 47: 249-260.

Scibelli, S., Shirley, Y. (2020) Prevalence of complex organic molecules in starless and prestellar cores within the Taurus molecular cloud. *Astrophys J* 891: 73.

Souza-Corrêa, J. A., da Costa, C. A. P., da Silveira, E. F. (2019). Compaction and destruction cross-sections for α-glycine from radiolysis process via 1.0 keV electron beam as a function of temperature. *Astrobiology* 19: 1123-1138.

Ursini, O., Angelini, G., Cataldo, F., Iglesias-Groth, S. (2019) Fullerene radiolysis in astrophysical ice analogs: a mass spectrometric study of the products. *Astrobiol* 19: 903-914.

Urey, H. C. (1955) The cosmic abundances of potassium, uranium, and thorium and the heat balances of the Earth, the Moon, and Mars. *Proc Nat Acad Sci* 41: 127-144.

Urey, H. C. (1956) The cosmic abundances of potassium, uranium, and thorium and the heat balances of the Earth, the Moon, and Mars. *Proc Nat Acad Sci* 42: 889-891.

Wolpert, M., Hellwig, P. (2006) Infrared spectra and molar absorption coefficients of the 20 alpha amino acids in aqueous solutions in the spectral range from 1800 to 500 cm$^{-1}$. *Spectrochim Acta Part A: Molec Biomolec Spectr* 64: 987-1001.

**TABLE 1 - MID- AND FAR-INFRARED BANDS OF GLYCINE WITH MOLAR EXTINCTION COEFFICIENTS AND INTEGRATED MOLAR ABSORPTIVITY**

| GLYCINE | | | | | λ in μm and T in Kelvin | L mol$^{-1}$ cm$^{-1}$ | km/mol |
|---|---|---|---|---|---|---|---|
| Wavenumber (cm$^{-1}$) | Wavelength (μm) | Integrated absorptivity | Integration region (cm$^{-1}$) | Absorbance | Bands with largest shift with temperature | **Epsilon** | **PSI** |
| 3176 | 3.15 | | 3702.7-2377.8 | 0.243 | | **15.0** | **-** |
| 3008 | 3.32 | | | | | | |
| 2974 | 3.36 | | | | | | |
| 2909 | 3.44 | 194.89 | | 0.279 | | **17.3** | **120.53** |
| 2829 | 3.53 | | | | | | |
| 2713 | 3.69 | | | 0.224 | | **13.9** | **-** |
| 2615 | 3.82 | | | 0.235 | | **14.5** | **-** |
| 2530 | 3.95 | | | 0.169 | | **10.5** | **-** |
| 2125 | 4.71 | 5.187 | 2252.4-2048.0 | 0.06 | | **3.7** | **3.21** |
| 1614 | 6.19 | | | 0.318 | | **19.7** | **-** |
| 1592 | 6.28 | 20.37 | 1662.3-1552.4 | 0.336 | | **20.8** | **12.60** |
| 1520 | 6.58 | 15.748 | 1554.3-1463.7 | 0.36 | | **22.3** | **9.74** |
| 1507 | 6.64 | | | 0.339 | | **21.0** | **-** |
| 1444 | 6.92 | 0.824 | 1456.0-1434.8 | 0.104 | | **6.4** | **0.51** |
| 1414 | 7.07 | 14.84 | 1436.7-1359.6 | 0.449 | | **27.8** | **9.18** |
| 1333 | 7.50 | 15.666 | 1361.5-1251.6 | 0.624 | | **38.6** | **9.69** |
| 1200 | 8.34 | | 1214.9-1180.2 | 0.006 | | **0.4** | **-** |
| 1132 | 8.83 | | | 0.16 | $\lambda = 2.900 \times 10^{-4} T + 8.743$ | **9.9** | **-** |
| 1113 | 8.99 | 6.845 | 1164.8-1074.2 | 0.212 | | **13.1** | **4.23** |
| 1034 | 9.67 | 3.098 | | 0.189 | | **11.7** | **1.92** |
| 910 | 10.99 | | | 0.501 | $\lambda = 3.995 \times 10^{-4} T + 10.835$ | **31.0** | **-** |
| 893 | 11.20 | 17.00 | 985.5-833.1 | 0.538 | | **33.3** | **10.51** |
| 698 | 14.32 | 10.587 | 759.8-646.0 | 0.413 | $\lambda = 5.445 \times 10^{-4} T + 14.151$ | **25.5** | **6.55** |
| 608 | 16.45 | 3.36 | 648.0-572.7 | 0.186 | $\lambda = 6.186 \times 10^{-4} T + 16.236$ | **11.5** | **2.08** |
| 504 | 19.86 | 20.484 | 586.2-445.2 | 0.524 | | **32.4** | **12.67** |
| *504* | *19.86* | *24.376* | *586.2-445.5* | *0.684* | | *42.3* | *15.08* |
| *359* | *27.85* | *23.325* | *445.5-293.1* | *0.631* | | *39.0* | *14.43* |
| *209* | *47.96* | *28.536* | *285.4-148.5* | *0.462* | | *28.6* | *17.65* |

| TABLE 2 - MID- AND FAR-INFRARED BANDS OF ISOLEUCINE WITH MOLAR EXTINCTION COEFFICIENTS AND INTEGRATED MOLAR ABSORPTIVITY | | | | | | | |
|---|---|---|---|---|---|---|---|
| **ISOLEUCINE** | | | | | $\lambda$ in µm and T in Kelvin | L mol$^{-1}$ cm$^{-1}$ | km/mol |
| Wavenumber (cm$^{-1}$) | Wavelength (µm) | Integrated absorptivity | Integration region (cm$^{-1}$) | Absorbance | Bands with largest shift with temperature or main bands dependence from temperature | **Epsilon** | **PSI** |
| 3059 | 3.27 | | | 0.458 | | **96.9** | - |
| 2970 | 3.37 | 250.212 | 3631.3-2408.6 | 0.690 | | **146.0** | **529.31** |
| 2879 | 3.47 | | | 0.449 | | **95.0** | - |
| 2745 | 3.64 | | | | | | |
| 2697 | 3.71 | | | | | | |
| 2623 | 3.81 | | | 0.21 | | **44.4** | - |
| 2116 | 4.73 | 5.614 | 2238.9-2019.1 | 0.06 | | **12.7** | **11.88** |
| 1608 | 6.22 | | | 0.671 | $\lambda = 8.944 \times 10^{-5}$ T + 6.192 | **141.9** | - |
| 1585 | 6.31 | 53.33 | 1685.5-1544.7 | 0.893 | $\lambda = -3.764 \times 10^{-5}$ T + 6.327 | **188.9** | **112.82** |
| 1514 | 6.61 | 23.73 | 1544.7-1479.1 | 1.080 | $\lambda = 4.465 \times 10^{-5}$ T + 6.593 | **228.5** | **50.20** |
| 1465 | 6.82 | 1.604 | 1479.1-1452.1 | 0.153 | | **32.4** | **3.39** |
| 1417 | 7.06 | | | 0.411 | | **86.9** | - |
| 1396 | 7.16 | 15.913 | 1440.6-1371.1 | 0.635 | $\lambda = 1.259 \times 10^{-4} T + 7.130$ | **134.3** | **33.66** |
| 1352 | 7.40 | 3.861 | 1371.1-1340.3 | 0.252 | | **53.3** | **8.17** |
| 1329 | 7.53 | 3.257 | 1340.3-1317.1 | 0.404 | | **85.5** | **6.89** |
| 1307 | 7.65 | 2.014 | 1317.1-1295.9 | 0.235 | | **49.7** | **4.26** |
| 1290 | 7.75 | | | 0.016 | | **3.3** | **-** |
| 1271 | 7.87 | | | 0.051 | | **10.7** | **-** |
| 1257 | 7.95 | | | 0.048 | | **10.2** | **-** |
| 1250 | 8.00 | | | 0.049 | | **10.4** | **-** |
| 1188 | 8.42 | 2.66 | 1234.2-1164.8 | 0.170 | | **36.0** | **5.63** |
| 1128 | 8.87 | 2.766 | 1164.8-1106.9 | 0.122 | $\lambda = 4.022 \times 10^{-4} T + 8.725$ | **25.8** | **5.85** |
| 1087 | 9.20 | 1.344 | 1106.9-1072.2 | 0.115 | | **24.3** | **2.84** |
| 1045 | 9.57 | 1.52 | 1072.2-1010.5 | 0.063 | | **13.3** | **3.22** |
| 1036 | 9.66 | | | 0.054 | | **11.4** | **-** |
| 993 | 10.07 | 0.626 | 1010.5-979.7 | 0.061 | | **12.9** | **1.32** |
| 964 | 10.38 | 0.86 | 979.7-948.8 | 0.082 | | **17.3** | **1.82** |
| 921 | 10.85 | 1.411 | 948.8-904.4 | 0.120 | | **25.4** | **2.98** |
| 873 | 11.45 | 1.525 | 904.4-858.2 | 0.122 | $\lambda = 2.7657 \times 10^{-4} T + 11.365$ | **25.8** | **3.23** |
| 852 | 11.73 | 0.134 | 858.2-840.8 | 0.027 | | **5.7** | **0.28** |
| 826 | 12.10 | 0.302 | 840.8-815.7 | 0.037 | | **7.8** | **0.64** |
| 802 | 12.47 | 0.526 | 815.7-792.6 | 0.075 | | **15.9** | **1.11** |
| 784 | 12.76 | 0.172 | 792.6-777.2 | 0.026 | | **5.5** | **0.36** |
| 769 | 13.00 | 0.489 | 777.2-759.8 | 0.070 | | **14.8** | **1.03** |
| 749 | 13.35 | 0.408 | 759.8-738.6 | 0.051 | | **10.8** | **0.86** |
| 712 | 14.05 | 2.53 | 738.6-696.2 | 0.278 | | **58.8** | **5.35** |
| 676 | 14.79 | 1.587 | 696.2-651.8 | 0.122 | | **25.8** | **3.36** |
| 557 | 17.94 | | | 0.194 | | **41.0** | **-** |
| 538 | 18.59 | 6.083 | 580.4-516.8 | 0.319 | $\lambda = 3.476 \times 10^{-4} T + 18.489$ | **67.5** | **12.87** |
| 496 | 20.18 | 0.714 | 516.8-474.4 | 0.034 | $\lambda = 5.535 \times 10^{-4} T + 19.990$ | **7.2** | **1.51** |
| 443 | 22.55 | 3.319 | 466.7-412.6 | 0.169 | | **35.8** | **7.02** |
| 426 | 23.46 | | | | | | |
| *557* | *17.94* | | | *0.131* | | *27.7* | *-* |
| *538* | *18.59* | *5.41* | *570.8-518.8* | *0.258* | | *54.6* | *11.44* |
| *496* | *20.18* | *0.654* | *518.8-476.3* | *0.025* | | *5.3* | *1.38* |
| *444* | *22.55* | *3.104* | *468.6-416.5* | *0.135* | | *28.6* | *6.57* |
| *426* | *23.46* | | | *0.017* | | *3.6* | *-* |
| *393* | *25.42* | *2.45* | *414.6-378.0* | *0.157* | | *33.2* | *5.18* |
| *370* | *27.01* | *0.248* | *378.0-360.6* | *0.029* | | *6.1* | *0.52* |
| *343* | *29.13* | *2.102* | *358.7-322.1* | *0.132* | | *27.9* | *4.45* |
| *314* | *31.81* | *0.458* | *322.1-300.8* | *0.047* | | *9.9* | *0.97* |
| *291* | *34.34* | | | *0.009* | | | |
| *280* | *35.76* | | | *0.006* | | | |
| *247* | *40.47* | *3.01* | *273.8-212.1* | *0.089* | | *18.8* | *6.37* |
| *227* | *44.11* | | | | | | |
| *202* | *49.39* | | | | | | |
| *172* | *58.26* | *0.63* | *196.7-162.0* | *0.027* | | *5.7* | *1.33* |

**TABLE 3 - MID- AND FAR-INFRARED BANDS OF PHENYLALANINE WITH MOLAR EXTINCTION COEFFICIENTS AND INTEGRATED MOLAR ABSORPTIVITY**

| PHENYLALANINE | | | | | $\lambda$ in µm and T in Kelvin | L mol$^{-1}$ cm$^{-1}$ | km/mol |
|---|---|---|---|---|---|---|---|
| Wavenumber (cm$^{-1}$) | Wavelength (µm) | Integrated absorptivity | Integration region (cm$^{-1}$) | Absorbance | Bands with largest shift with temperature or main bands dependence from temperature | Epsilon | PSI |
| 3446 | 2.90 | | 3693.0-2231.2 | 0.155 | | **27.5** | |
| 3068 | 3.26 | 239.73 | | 0.488 | | **86.6** | **425.57** |
| 2964 | 3.37 | | | 0.399 | | **70.8** | |
| 2856 | 3.50 | | | 0.179 | | **31.8** | |
| 2737 | 3.65 | | | 0.137 | | **24.3** | |
| 2551 | 3.92 | | | 0.155 | | **27.5** | |
| 2430 | 4.12 | | | 0.149 | | **26.5** | |
| 2129 | 4.70 | 12.914 | 2231.2-1990.2 | 0.132 | | **23.4** | **22.93** |
| 1626 | 6.15 | | 1822.4-1429.0 | 0.558 | $\lambda = 7.040 \times 10^{-5}\,T + 6.125$ | **99.1** | |
| 1603 | 6.24 | | | 0.581 | | **103.1** | |
| 1564 | 6.39 | 125.51 | | 1.419 | $\lambda = 9.373 \times 10^{-6}\,T + 6.400$ | **251.9** | **222.81** |
| 1496 | 6.69 | | | 0.648 | | **115.0** | |
| 1458 | 6.86 | | | 0.185 | $\lambda = -1.086 \times 10^{-5}\,T + 6.694$ | **32.8** | |
| 1446 | 6.91 | | | 0.137 | | **24.3** | |
| 1412 | 7.08 | 11.41 | 1430.9-1373.1 | 0.744 | $\lambda = 8.031 \times 10^{-5}\,T + 7.066$ | **132.1** | **20.26** |
| 1340 | 7.46 | | 1371.1-1265.1 | 0.368 | | **65.3** | - |
| 1321 | 7.57 | | | 0.431 | | **76.5** | - |
| 1307 | 7.65 | 24.85 | | 0.830 | $\lambda = 7.006 \times 10^{-5}\,T + 7.628$ | **147.3** | **44.11** |
| 1294 | 7.73 | | | 0.322 | | **57.2** | - |
| 1226 | 8.15 | 2.557 | 1259.3-1195.6 | 0.134 | | **23.8** | **4.54** |
| 1165 | 8.59 | 4.801 | 1195.6-1141.6 | 0.243 | | **43.1** | **8.52** |
| 1130 | 8.85 | 0.875 | 1141.6-1106.9 | 0.068 | | **12.1** | **1.55** |
| 1074 | 9.31 | 1.941 | 1106.9-1058.7 | 0.197 | | **35.0** | **3.45** |
| 1026 | 9.75 | 1.926 | 1054.9-1014.4 | 0.110 | | **19.5** | **3.42** |
| 1005 | 9.95 | 1.200 | 1014.4-976.7 | 0.092 | | **16.3** | **2.13** |
| 951 | 10.52 | 0.977 | 964.2-933.4 | 0.071 | | **12.6** | **1.73** |
| 914 | 10.94 | 0.942 | 933.4-896.7 | 0.099 | | **17.6** | **1.67** |
| 950 | 10.52 | 4.076 | 896.7-823.4 | 0.293 | | **52.0** | **7.24** |
| 779 | 12.84 | 1.823 | 804.2-767.5 | 0.190 | | **33.7** | **3.24** |
| 746 | 13.40 | 5.164 | 767.5-730.9 | 0.518 | | **92.0** | **9.17** |
| 700 | 14.29 | 11.166 | 730.9-647.9 | 0.786 | | **139.5** | **19.82** |
| 683 | 14.65 | | | 0.208 | | **36.9** | - |
| 605 | 16.51 | 1.172 | 640.2-566.9 | 0.102 | | **18.1** | **2.08** |
| 526 | 18.99 | 7.862 | 566.9-499.5 | 0.610 | | **108.3** | **13.96** |
| 469 | 21.34 | 2.889 | 499.5-447.4 | 0.197 | | **35.0** | **5.13** |
| *526* | *18.99* | *8.726* | *590.1-503.3* | *0.685* | | *121.6* | *15.49* |
| *469* | *21.34* | *2.742* | *497.5-445.5* | *0.213* | | *37.8* | *4.87* |
| *366* | *27.31* | *30.406* | *445.5-287.3* | *0.673* | | *119.5* | *53.98* |
| *214* | *46.72* | | *237.2-169.7* | *0.103* | | *18.3* | - |
| *197* | *50.84* | *5.15* | *237.2-169.7* | *0.17* | | *30.2* | *9.14* |

| TABLE 4 - MID- AND FAR-INFRARED BANDS OF TYROSINE WITH MOLAR EXTINCTION COEFFICIENTS AND INTEGRATED MOLAR ABSORPTIVITY | | | | | | | |
|---|---|---|---|---|---|---|---|
| **TYROSINE** | | | | | $\lambda$ in µm and T in Kelvin | **L mol$^{-1}$ cm$^{-1}$** | **km/mol** |
| Wavenumber (cm$^{-1}$) | Wavelength (µm) | Integrated absorptivity | Integration region (cm$^{-1}$) | Absorbance | Bands with largest shift with temperature or main bands dependence from temperature | **Epsilon** | **PSI** |
| 3208 | 3.12 | 312.81 | 3382.5-2231.3 | 0.794 | | **187.2** | **737.32** |
| 2128 | 4.70 | | | 0.442 | | **104.2** | **-** |
| 2960 | 3.38 | | | 0.635 | | **149.7** | **-** |
| 2931 | 3.41 | | | 0.604 | | **142.4** | **-** |
| 2883 | 3.47 | | | 0.481 | | **113.4** | **-** |
| 2829 | 3.53 | | | 0.442 | | **104.2** | **-** |
| 2755 | 3.63 | | | 0.382 | | **90.0** | **-** |
| 2648 | 3.78 | | | 0.296 | | **69.8** | **-** |
| 2600 | 3.85 | | | 0.293 | | **69.1** | **-** |
| 2505 | 3.99 | | | | | | |
| 2484 | 4.03 | | | 0.117 | | **27.6** | **-** |
| 2083 | 4.80 | 5.311 | 2183.0-1996.0 | 0.081 | | **19.1** | **12.52** |
| 1608 | 6.22 | | 1718.3-1540.8 | 1.593 | $\lambda = 7.632 \times 10^{-5}$ T + 6.181 | **375.5** | **-** |
| 1589 | 6.29 | 78.786 | | 1.649 | $\lambda = 4.922 \times 10^{-5}$ T + 6.271 | **388.7** | **185.71** |
| 1514 | 6.61 | 11.904 | 1537.0-1488.8 | 0.743 | $\lambda = 2.653 \times 10^{-5}$ T + 6.604 | **175.1** | **28.06** |
| 1454 | 6.88 | 17.622 | 1488.8-1396.2 | 0.49 | $\lambda = 7.799 \times 10^{-5}$ T + 6.853 | **115.5** | **41.54** |
| 1435 | 6.97 | | | 0.306 | | **72.1** | **-** |
| 1417 | 7.06 | | | 0.39 | | **91.9** | **-** |
| 1377 | 7.26 | 12.042 | 1396.2.-1348.0 | 0.297 | | **70.0** | **28.38** |
| 1363 | 7.33 | | | 0.754 | $\lambda = 3.836 \times 10^{-5}\,T + 7.327$ | **177.7** | - |
| 1331 | 7.52 | 15.015 | 1347.1-1292.1 | 1.314 | $\lambda = 5.609 \times 10^{-5}\,T + 7.499$ | **309.7** | **35.39** |
| 1267 | 7.89 | | 1292.1-1203.4 | 0.279 | | **65.8** | - |
| 1245 | 8,03 | 25.251 | | 1.310 | $\lambda = 1.535 \times 10^{-4}\,T + 7.985$ | **308.8** | **59.52** |
| 1215 | 8.23 | | | 0.211 | | **49.7** | - |
| 1198 | 8.35 | 0.529 | 1203.4-1189.9 | 0.091 | | **21.4** | **1.25** |
| 1176 | 8.50 | | 1189.9-1132.0 | 0.226 | | **53.3** | - |
| 1155 | 8.66 | 4.452 | | 0.189 | | **44.5** | **10.49** |
| 1113 | 8.99 | | 1132.0-1080.0 | 0.205 | | **48.3** | - |
| 1099 | 9.10 | 3.851 | | 0.242 | | **57.0** | **9.08** |
| 1042 | 9.59 | 2.364 | 1080.0-1022.1 | 0.235 | | **55.4** | **5.57** |
| 985 | 10.15 | 1.171 | 1004.7-952.7 | 0.112 | | **26.4** | **2.76** |
| 939 | 10.65 | 0.333 | 952.7-927.6 | 0.053 | | **12.5** | **0.78** |
| 897 | 11.15 | 0.953 | 914.1-887.1 | 0.122 | | **28.8** | **2.25** |
| 877 | 11.40 | 1.963 | 889.0-862.0 | 0.234 | | **55.2** | **4.63** |
| 841 | 11.89 | 6.150 | 862.0-821.5 | 0.601 | | **141.7** | **14.50** |
| 802 | 12.46 | | | | | | |
| 794 | 12.59 | 5.386 | 821.5-761.6 | 0.22 | | **51.9** | **12.70** |
| 740 | 13.50 | 2.088 | 761.6-723.2 | 0.262 | | **61.8** | **4.92** |
| 713 | 14.01 | 0.500 | 723.2-690.4 | 0.079 | | **18.6** | **1.18** |
| 650 | 15.38 | 5.384 | 682.7-605.5 | 0.396 | | **93.3** | **12.69** |
| 576 | 17.36 | 5.397 | 605.5-559.2 | 0.577 | | **136.0** | **12.72** |
| 530 | 18.86 | 6.803 | 559.2-503.3 | 0.465 | | **109.6** | **16.04** |
| 494 | 20.26 | 1.16 | 503.3-482.1 | 0.167 | | **39.4** | **2.73** |
| 474 | 21.08 | 0.268 | 484.0-462.8 | 0.026 | | **6.1** | **0.63** |
| 451 | 22.16 | | | | | | |
| 434 | 23.05 | 2.339 | 459.0- | 0.146 | | **34.4** | **5.51** |

| | | | 416.5 | | | | |
|---|---|---|---|---|---|---|---|
| *576* | *17.36* | *5.397* | *597.8-559.3* | *0.623* | | *146.8* | *12.72* |
| *530* | *18.86* | *7.199* | *559.3-505.3* | *0.509* | | *120.0* | *16.97* |
| *494* | *20.26* | *1.162* | *505.3-484.0* | *0.177* | | *41.7* | *2.74* |
| *474* | *21.08* | *0.268* | *484.0-462.8* | *0.026* | | *6.1* | *0.63* |
| *451* | *22.16* | | *484,0-462.8* | *0.041* | | *9.7* | |
| *434* | *23.05* | *2.788* | *462.8-410.8* | *0.190* | | *44.8* | *6.57* |
| *380* | *26.32* | *6.617* | *410.8-360.2* | *0.461* | | *108.7* | *15.60* |
| *335* | *29.80* | *0.565* | *345.2-327.8* | *0.075* | | *17.7* | *1.33* |
| *311* | *32.10* | *3.001* | *327.8-291.2* | *0.250* | | *58.9* | *7.07* |
| *281* | *35.52* | | | *0.010* | | *2.4* | |
| *247* | *40.47* | *6.107* | *273.8-231.4* | *0.508* | | *119.7* | *14.39* |
| *208* | *48.01* | *2.409* | *231.4-189.0* | *0.149* | | *35.1* | *5.68* |
| *170* | *58.92* | | | *0.021* | | *4.9* | |
| *156* | *64.02* | *1.202* | *167.8-146.6* | *0.100* | | *23.6* | *2.83* |

| TABLE 5 - MID- AND FAR-INFRARED BANDS OF TRIPTOPHAN WITH MOLAR EXTINCTION COEFFICIENTS AND INTEGRATED MOLAR ABSORPTIVITY | | | | | | | |
|---|---|---|---|---|---|---|---|
| **TRYPTOPHAN** | | | | | $\lambda$ in µm and T in Kelvin | **L mol$^{-1}$ cm$^{-1}$** | **km/mol** |
| Wavenumber (cm$^{-1}$) | Wavelength (µm) | Integrated absorptivity | Integration region (cm$^{-1}$) | Absorbance | Bands with largest shift with temperature or main bands dependence from temperature | **Epsilon** | **PSI** |
| 3404 | 2.94 | 27.042 | 3478.9-3353.6 | 0.924 | | **259.1** | **75.84** |
| 3080 | 3.25 | | 3343.9-2171.4 | 0.589 | | **165.2** | **-** |
| 3048 | 3.28 | 212.3 | | 0.588 | | **164.9** | **595.37** |
| 2854 | 3.50 | | | 0.210 | | **58.9** | **-** |
| 2735 | 3.66 | | | 0.207 | | **58.1** | **-** |
| 2567 | 3.90 | | | 0.226 | | **63.4** | **-** |
| 2077 | 4.81 | 5.299 | 2169.5-1982.5 | 0.065 | | **18.2** | **14.86** |
| 1666 | 6.00 | 92.06 | 1789.6-1494.4 | 1.175 | | **329.5** | **258.17** |
| 1593 | 6.28 | | | 1.015 | | **284.6** | **-** |
| 1487 | 6.72 | 0.27 | 1494.6-1479.1 | 0.047 | | **13.2** | **0.76** |
| 1458 | 6.86 | 4.282 | 1479.1-1440.6 | 0.480 | $\lambda = 2.869 \times 10^{-5}$ T + 6.854 | **134.6** | **12.01** |
| 1415 | 7.06 | 17.63 | 1442.5-1376.9 | 0.805 | $\lambda = 1.629 \times 10^{-4}$ T + 7.018 | **225.8** | **49.44** |
| 1358 | 7.37 | 13.21 | 1376.9-1324.9 | 0.884 | $\lambda = 5.784 \times 10^{-5}$ T + 7.349 | **247.9** | **37.05** |
| 1316 | 7.60 | 0.965 | 1324.9-1307.5 | 0.137 | | **38.4** | **2.71** |
| 1294 | 7.73 | 0.287 | 1307.5-1286.3 | 0.036 | | **10.1** | **0.80** |
| 1279 | 7.82 | 0.311 | 1286.3-1267.0 | 0.034 | | **9.5** | **0.87** |
| 1252 | 7.99 | 0.393 | 1267.0-1240.0 | 0.041 | | **11.5** | **1.10** |
| 1231 | 8.12 | 1.267 | 1240.0- | 0.168 | | **47.1** | **3.55** |

| | | | 1214.9 | | | | |
|---|---|---|---|---|---|---|---|
| 1186 | 8.43 | | | 0.134 | | **37.6** | **-** |
| 1158 | 8.64 | 3.035 | 1186.0-1135.9 | 0.169 | | **47.4** | **8.51** |
| 1132 | 8.83 | 0.136 | 1135.9-1124.3 | 0.020 | | **5.6** | **0.38** |
| 1115 | 8.97 | 0.125 | 1124.3-1108.9 | 0.032 | | **9.0** | **0.35** |
| 1101 | 9.09 | 1.635 | 1110.8-1083.8 | 0.145 | | **40.7** | **4.59** |
| 1076 | 9.29 | 0.416 | 1083.8-1070.3 | 0.069 | | **19.4** | **1.17** |
| 1053 | 9.50 | 0.276 | 1064.5-1043.3 | 0.031 | | **8.7** | **0.77** |
| 1009 | 9.91 | 1.375 | 1043.3-997.2 | 0.122 | | **34.2** | **3.86** |
| 987 | 10.13 | 0.956 | 997.2-968.1 | 0.095 | | **26.6** | **2.68** |
| 920 | 10.87 | 1.569 | 952.7-902.5 | 0.120 | | **33.7** | **4.40** |
| 865 | 11.56 | 3.852 | 890.8-823.4 | 0,146 | | **40.9** | **10.80** |
| 804 | 12.44 | 0.415 | 821.5-790.7 | 0,049 | | **13.7** | **1.16** |
| 767 | 13.03 | 0.262 | 775.2-761.7 | 0,055 | | **15.4** | **0.73** |
| 744 | 13.43 | 10.892 | 761.3-721.2 | 1,153 | $\lambda = 2.676 \times 10^{-3} T + 13.013$ | **323.3** | **30.55** |
| 708 | 14.13 | 0.138 | 721.2-696.2 | 0,017 | | **4.8** | **0.39** |
| 683 | 14.65 | 0.813 | 696.2-667.2 | 0,081 | | **22.7** | **2.28** |
| 656 | 15.25 | 0.166 | 667.2-647.9 | 0,019 | $\lambda = 3.333 \times 10^{-4} T + 15.133$ | **5.3** | **0.47** |
| 627 | 15.96 | 0.968 | 647.9-611.3 | 0,111 | $\lambda = 3.007 \times 10^{-4} + 15.881$ | **31.1** | **2.71** |
| 588 | 17.00 | | 611.3-574.7 | 0,119 | | **33.4** | **-** |
| 580 | 17.23 | 1.989 | | 0,137 | | **38.4** | **5.58** |
| 559 | 17.88 | 3.483 | 574.7-541.9 | 0,188 | $\lambda = 6.593 \times 10^{-4} T + 17.715$ | **52.7** | **9.77** |
| 550 | 18.19 | | | 0,187 | | **52.4** | **-** |
| 528 | 18.92 | | 541.9-472.5 | 0,3 | $\lambda = 4.335 \times 10^{-4} T + 18.819$ | **84.1** | **-** |
| 509 | 19.64 | 11.308 | | 0,331 | $\lambda = 6.091 \times 10^{-4} + 19.497$ | **92.8** | **31.71** |
| 499 | 20.02 | | | 0,254 | $\lambda = 3.804 \times 10^{-4} T + 19.926$ | **71.2** | **-** |
| 457 | 21.88 | 0.665 | 470.5-447.4 | 0,094 | | **26.4** | **1.86** |
| 426 | 23.46 | 3.709 | 447.4-408.3 | 0,39 | | **109.4** | **10.40** |
| *588* | *17.00* | | *599.8-576.6* | *0,112* | | *31.4* | *-* |
| *580* | *17.23* | *1.711* | *599.8-576.6* | *0,145* | | *40.7* | *4.80* |
| *559* | *17.88* | *3.609* | *576.6-541.9* | *0,193* | | *54.1* | *10.12* |
| *550* | *18.19* | | *576.6-541.9* | *0,189* | | *53.0* | *-* |
| *528* | *18.92* | | *541.9-468.6* | *0,225* | | *63.1* | *-* |
| *509* | *19.64* | *11.400* | *541.9-468.6* | *0,242* | | *67.9* | *31.97* |
| *499* | *20.02* | | *541.9-468.6* | *0,194* | | *54.4* | *-* |
| *457* | *21.88* | *0.678* | *468.6-447.4* | *0,095* | | *26.6* | *1.90* |
| *426* | *23.46* | *3.551* | *447.4-408.8* | *0,401* | | *112.5* | *9.96* |
| *397* | *25.17* | *0.211* | *408.8-389.6* | *0,031* | | *8.7* | *0.59* |
| *357* | *28.03* | | *389.6-335.6* | *0,143* | | *40.1* | *-* |
| *349* | *28.65* | *3.410* | *389.6-335.6* | *0,169* | | *47.4* | *9.56* |
| *324* | *30.85* | *1.14* | *335.6-293.1* | *0,086* | | *24.1* | *3.20* |
| *269* | *37.10* | *0.647* | *285.4-260.4* | *0,067* | | *18.8* | *1.81* |
| *253* | *39.52* | *0.253* | *260.3-244.9* | *0,038* | | *10.7* | *0.71* |
| *237* | *42.16* | *0.612* | *244.9-229.5* | *0,085* | | *23.8* | *1.72* |
| *215* | *46.54* | | *229.5-179.3* | *0,108* | | *30.3* | *-* |
| *197* | *50.84* | *3.682* | *229.5-179.3* | *0,143* | | *40.1* | *10.33* |
| *170* | *58.92* | | | *0,01* | | *2.8* | *-* |
| *158* | *63.24* | | | *0,014* | | *3.9* | *-* |
| *150* | *66.48* | | | *0.007* | | *2.0* | *-* |

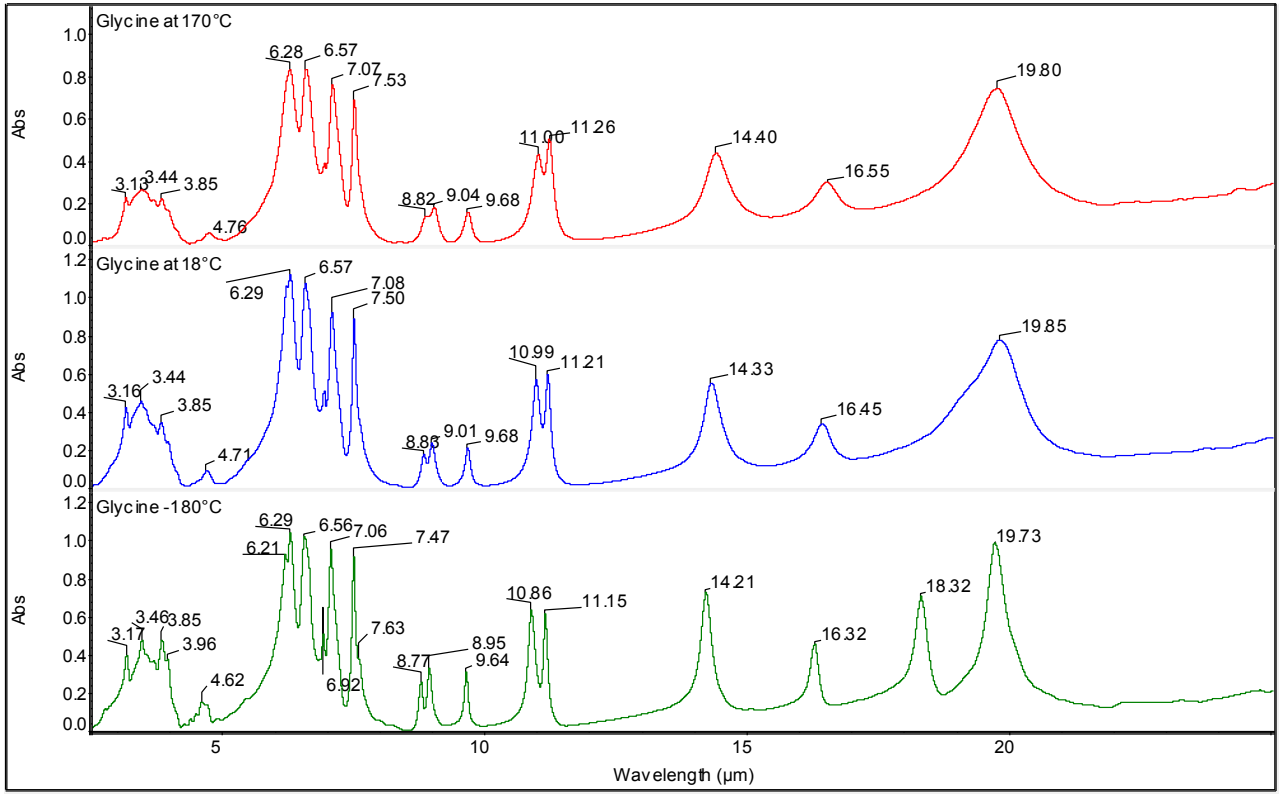

Fig. 1 – FT-IR spectra of glycine in CsI matrix recorded (from top to bottom) at +170°C, +18°C and -180°C respectively.

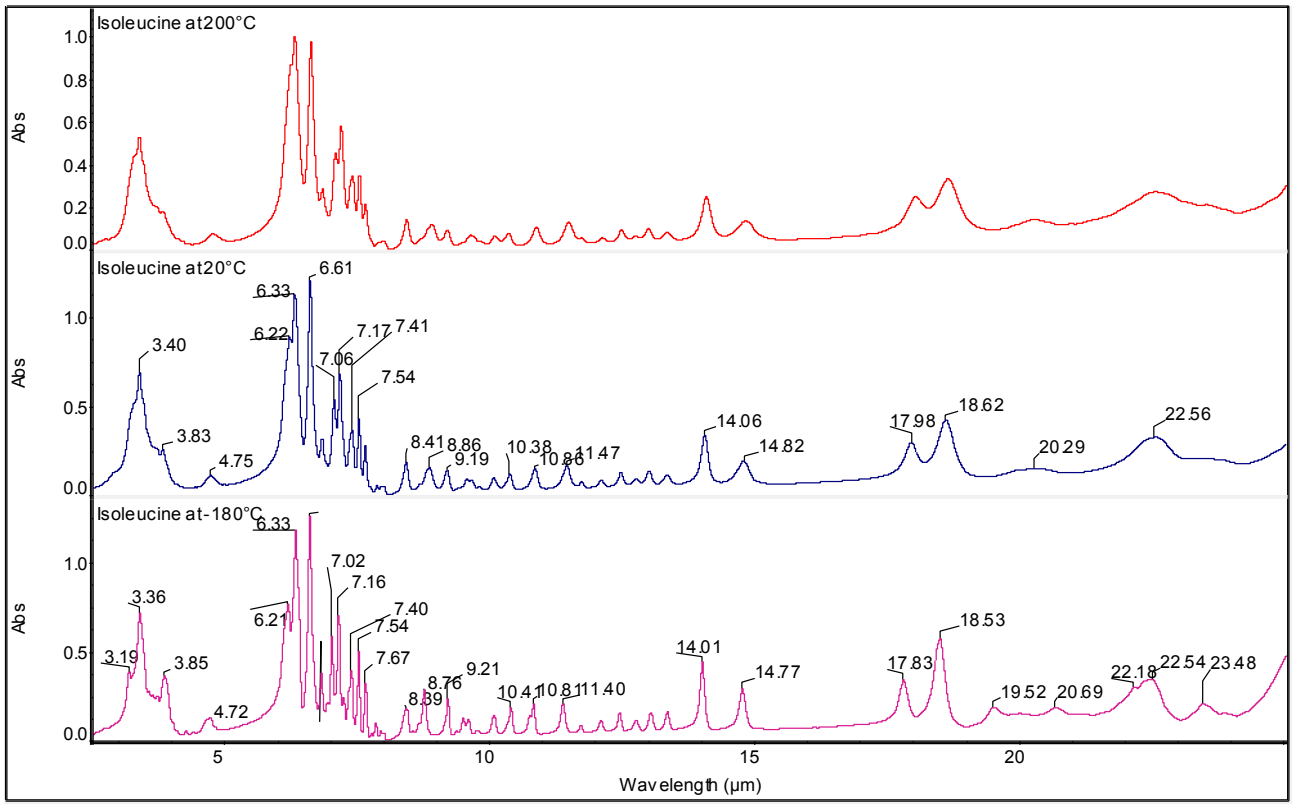

Fig. 2 – FT-IR spectra of isoleucine in CsI matrix recorded (from top to bottom) at +200°C, +20°C and -180°C respectively.

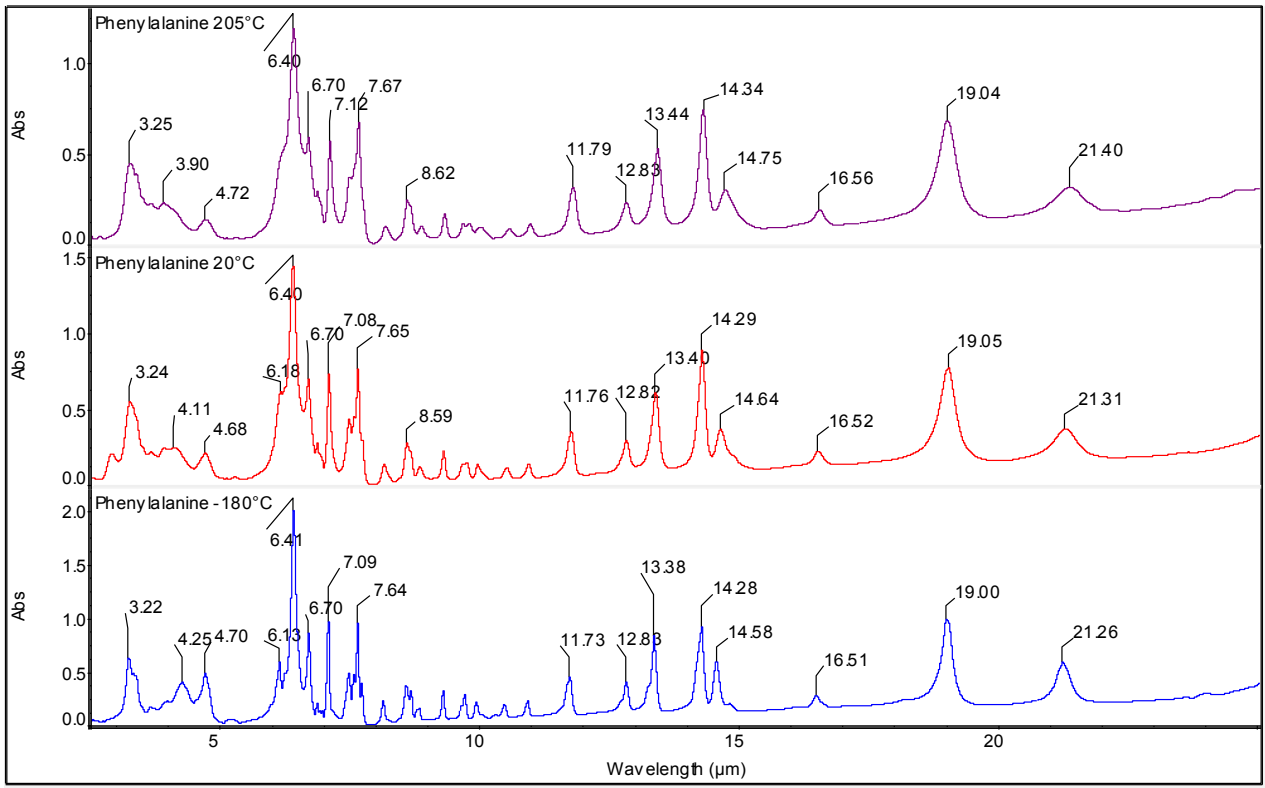

Fig. 3 – FT-IR spectra of phenylalanine in CsI matrix recorded (from top to bottom) at +205°C, +20°C and -180°C respectively.

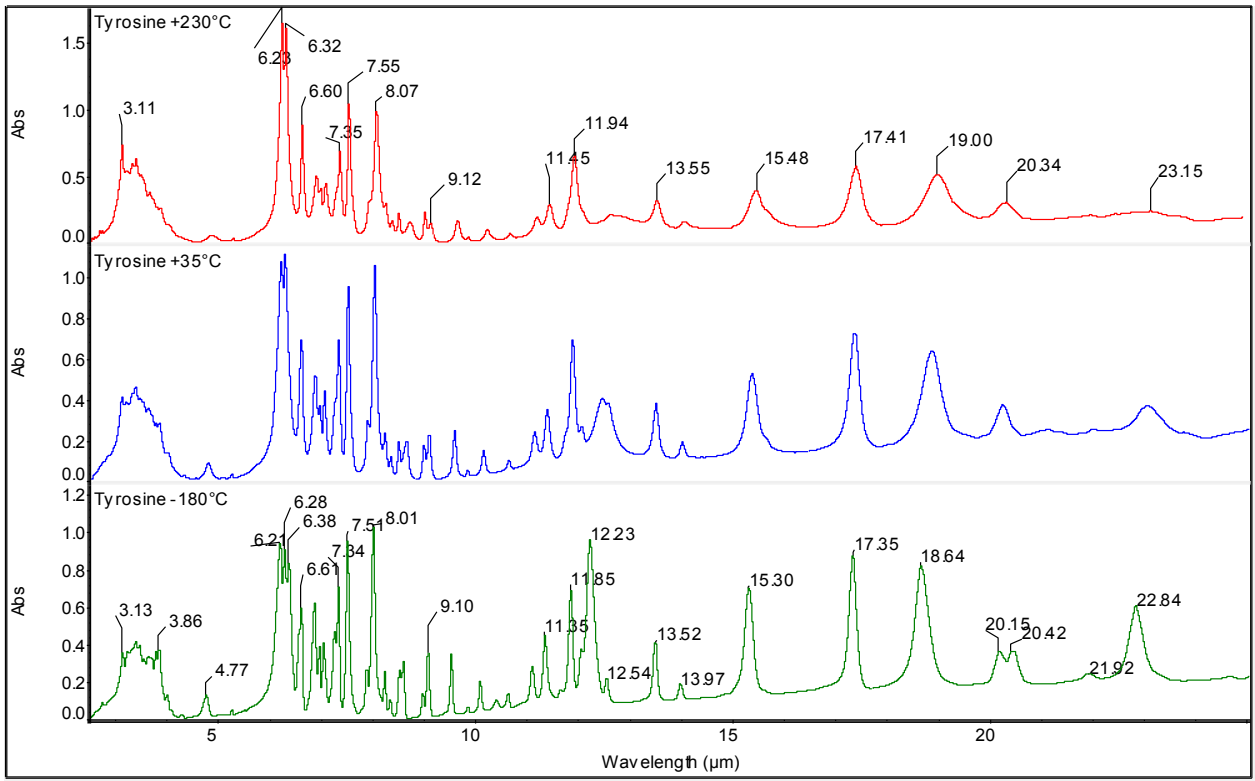

Fig. 4 – FT-IR spectra of tyrosine in CsI matrix recorded (from top to bottom) at +230°C, +35°C and -180°C respectively.

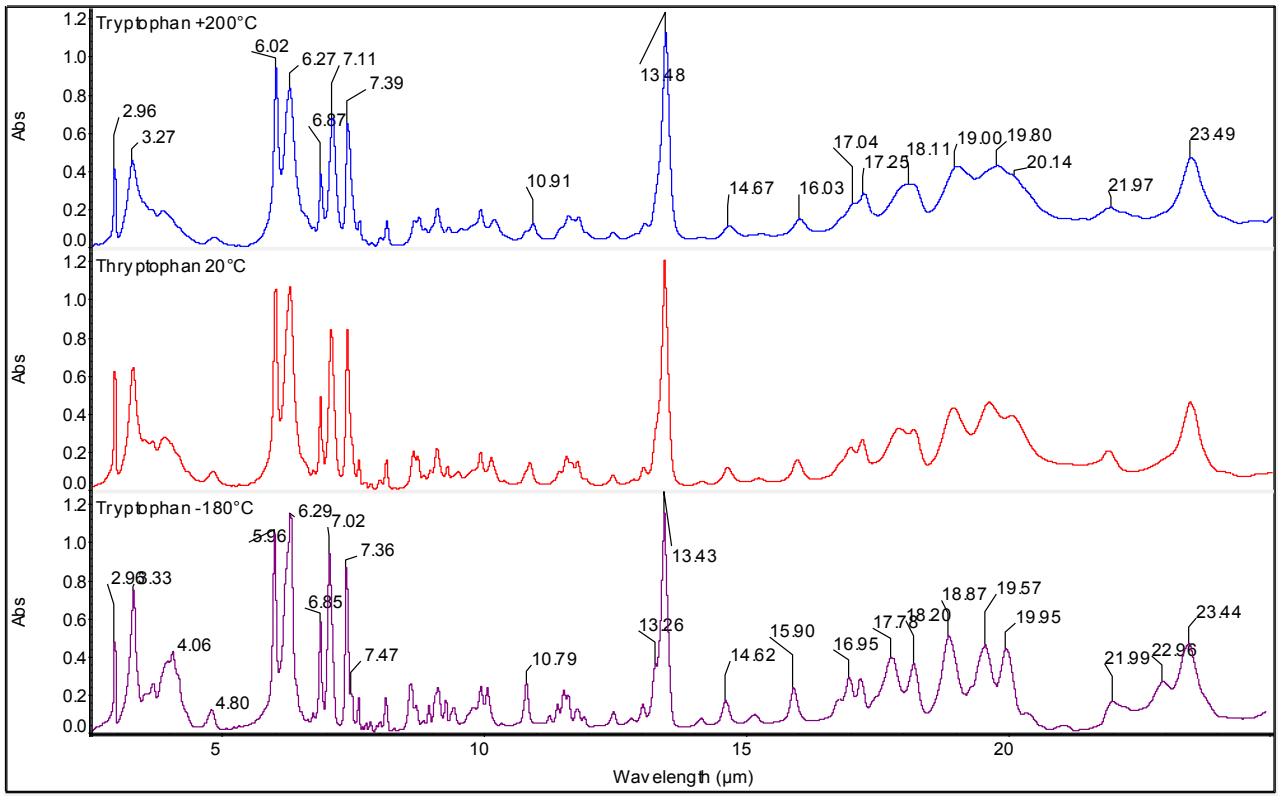

Fig. 5 – FT-IR spectra of tryptophan in CsI matrix recorded (from top to bottom) at +200°C, +20°C and -180°C respectively.

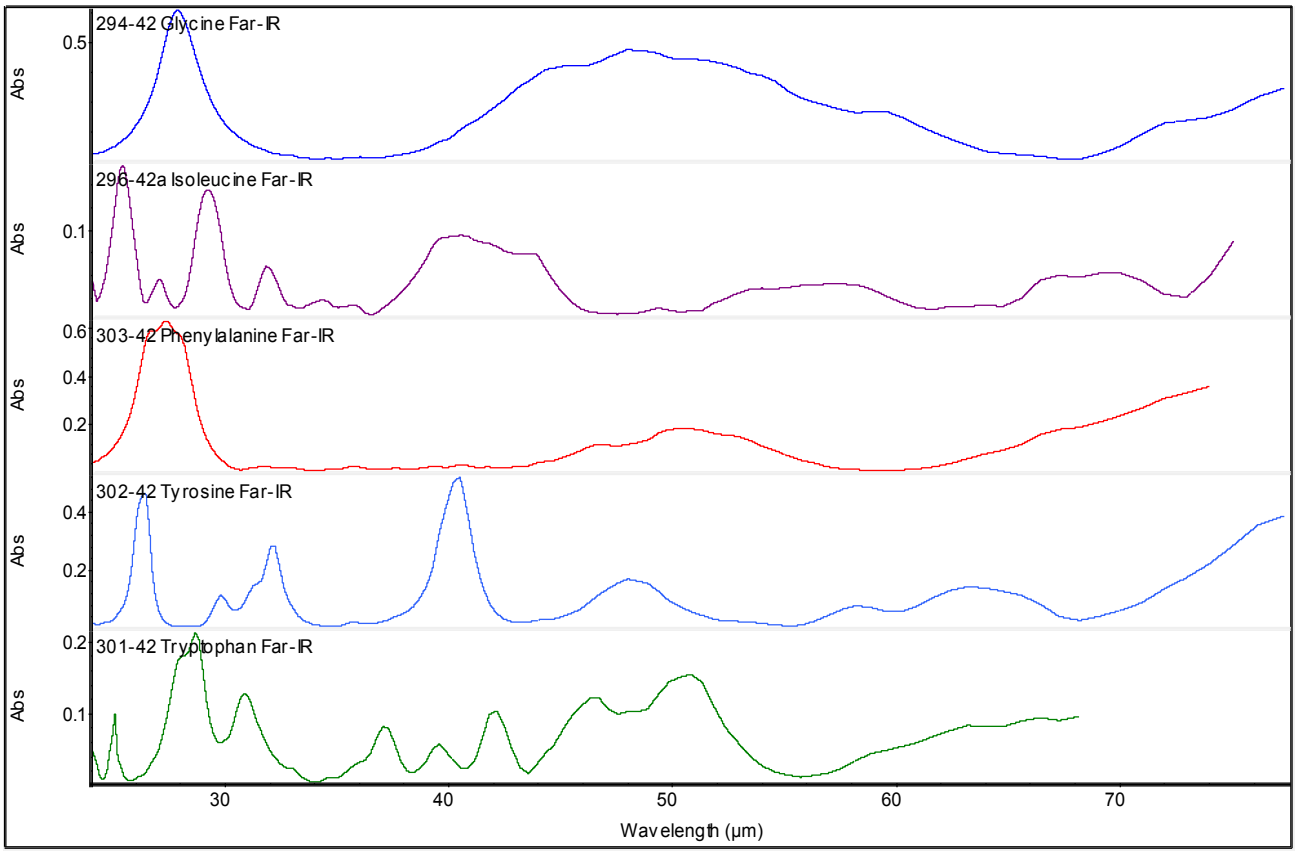

Fig. 6 – FT-IR spectra in the far-infrared range recorded at ambient temperature on CsI matrix; from top to bottom: glycine, isoleucine, phenylalanine, tyrosine and tryptophan.